\documentclass[conference]{IEEEtran}
\usepackage{amsmath,amsfonts}
\usepackage{graphicx}
\usepackage{enumitem}

\usepackage{algorithm}
\usepackage{bm}
\usepackage{array}
\usepackage{mathtools}
\usepackage{color,soul}
\usepackage{siunitx}
\usepackage[table,xcdraw]{xcolor}

\def\BibTeX{{\rm B\kern-.05em{\sc i\kern-.025em b}\kern-.08em
T\kern-.1667em\lower.7ex\hbox{E}\kern-.125emX}}
\usepackage{balance}
\usepackage{multirow}
\usepackage{cite}

\usepackage[utf8]{inputenc}

\usepackage{pgfplots}
\usepackage{breqn}
\usepackage[inkscapeformat=png]{svg}
\usepackage{stackengine}

\pgfplotsset{compat=1.14}

\makeatletter
\newcommand{\@giventhatstar}[2]{\left[#1\;\middle|\;#2\right]}
\newcommand{\@giventhatnostar}[3][]{#1[#2\;#1|\;#3#1]}
\newcommand{\giventhat}{\@ifstar\@giventhatstar\@giventhatnostar}

\makeatother

\begin{document}
\title{Filter-Banks for Ultra-Wideband for Communications, Sensing, and Localization
}
\author{Brian Nelson, Hussein Moradi, and Behrouz Farhang-Boroujeny}

\maketitle

\copyright 2025 IEEE. Personal use of this material is permitted. Permission from IEEE must be obtained for all other uses, in any current or future media, including reprinting/republishing this material for advertising or promotional purposes, creating new collective works, for resale or redistribution to servers or lists, or reuse of any copyrighted component of this work in other works.

\begin{abstract}
  Recently, filter-bank multicarrier spread spectrum (FBMC-SS) has been proposed as a candidate waveform for ultra-wideband (UWB) communications, sensing, and localization. It has been noted that FBMC-SS is a perfect match to this application, leading to a trivial method of matching to the required spectral mask at different regions of the world. FBMC-SS also allows easy rejection of high-power interfering signals that may appear over different parts of the UWB spectral band. Moreover, passing the received signal through a matched filter provides precise information for sensing and localization.  In this paper, we concentrate on the use of staggered multitone spread spectrum (SMT-SS) for UWB communications. SMT makes use of offset quadrature amplitude modulation (OQAM) to transmit data symbols over overlapping subcarrier bands.  This form of FBMC-SS is well-suited to UWB communications because it has good spectral efficiency and a flat power spectral density (PSD), resulting in good utilization of the UWB spectral mask.
\end{abstract}

\section{Introduction} \label{sec:intro}
The idea of ultra-wideband (UWB) communications may be traced back to the spark gap radio transmission methods that were invented towards the end of the nineteenth century and the beginning of the twentieth century by Marconi and others. Spark gap radios, also known as impulse radios, were used for wireless telegraphy at the time. In the 1990s, when wireless technologies were booming and spectrum availability was becoming scarce, the possibility of using impulse radio UWB (IR-UWB) to develop spread spectrum radios that could coexist with existing (relatively, narrowband) radios with a minimum/acceptable interference to them was brought up, \cite{Win1998,Win2000}. The idea was well received by the Federal Communications Commission (FCC) which in 2002 allocated 7.5 GHz of spectrum for unlicensed use of UWB devices in the 3.1 to 10.6 GHz frequency band. This was to allow communications over short distances of up to 10 to 20 meters without licensing the spectrum.

\begin{figure}
  \includegraphics[width=0.95\columnwidth]{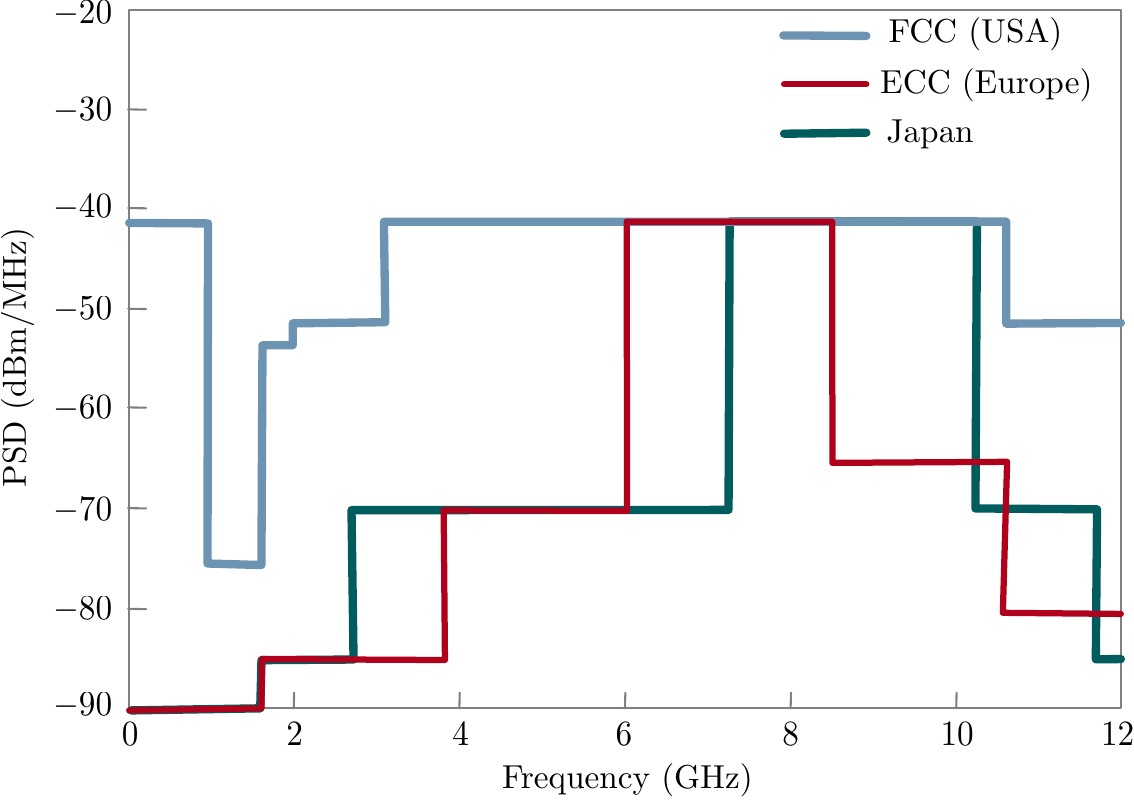}
  \caption{UWB spectral mask in USA, Europe, and Japan. }
  \label{fig:SpectralMask}
\end{figure}

The FCC had no waveform recommendation for UWB communications and instead specified that a UWB signal must occupy at least 500 MHz of bandwidth or have a bandwidth equivalent of at least 20\% of its center frequency, while satisfying a certain spectral mask for the transmit power spectral density (PSD) \cite{UWB2024}. This spectral mask is defined by the regulatory bodies and varies in different regions of the world.
Fig.~\ref{fig:SpectralMask} presents the spectral masks that have been defined for the USA (FCC), Europe (ECC), and Japan. As seen, these masks vary widely from place to place, hence, to be useable around the globe, a great degree of flexibility has to be included in any UWB system design.

After release of the 7.5 GHz of spectrum by the FCC, the two industrial organizations that emerged to establish standards for UWB, following IEEE802.15.3a, decided on waveforms that were not based on impulse radio. The first organization, UWB Forum, adopted a direct sequence (DS) spread spectrum-based UWB that was then referred to as DS-UWB, \cite{GiffordIan2004CtOU}.  The second organization, the WiMedia Alliance, advocated for multiband orthogonal frequency division multiplexing (MB-OFDM) \cite{MB-OFDM2004b}. Despite significant efforts and investments by a large body of wireless industry, developments under IEEE802.15.3a had effectively stalled in 2006, due to standardization issues. An excellent history summarizing the first few years after the FCC allocation of the 7.5 GHz of spectrum for UWB communications can be found in \cite{UWB2015}.

The subsequent follow up by the industry led to the IEEE802.15.4 series of standards, \cite{IEEE802.15.4a2010,IEEE802.15.4z2019}. The data rates in IEEE802.15.4 standards are significantly lower than the original rates that were envisioned in IEEE802.15.3a (up to 480 Mb/s). As opposed to IEEE802.15.3, IEEE802.15.4 standards adopted IR-UWB.

In a recent article, \cite{UWB2024}, filter-bank multicarrier (FBMC) has been introduced as a new candidate waveform for UWB communications. It has been noted that FBMC is a perfect match to the needs of UWB as it offers the following advantages:
\begin{itemize}
  \item
        Spectral mask adoption in different regions of the world is simplified through activation and deactivation of subcarriers.
  \item
        The filter-bank structure lends itself to implementations that are tailored for parallel processing and avoid power-hungry high-speed analog-to-digital (A/D) and digital-to-analog (D/A) convertors.
  \item
        The filter-bank structure allows adoption of a special receiver structure that blindly removes any interference. No other waveform in the broad area of digital communications offers this feature.
  \item
        FB-UWB design supporting simultaneous operation of piconets is straightforward.
  \item
        Sensing and ranging capability of FB-UWB is as straightforward as that of the IR-UWB.
\end{itemize}

The article  \cite{UWB2024} presents some numerical results showing FBMC based UWB can support transmission rates comparable to the best envisioned rates in the past literature.  These numerical results are for the cases where either filtered multitone (FMT) or staggered multi-tone (SMT),  \cite{OFDMvsFBMC}, are used to implement an FBMC spread spectrum (FBMC-SS) waveform. However, details of any transceiver design has been left for future studies.

In this paper, we explore a transceiver design based on SMT. SMT, in contrast to FMT, has a flat PSD and more efficiently utilizes the available spectrum. This is because the FMT subcarrier bands, by design, are non-overlapping. This leaves a null band between each pair of adjacent subcarrier bands. The presence of these spectral nulls, along with the need of satisfying the constraint imposed by the spectral mask, reduces the maximum power of the transmit signal, hence, leading to a reduced transmission range. On the other hand, by making use of offset quadrature amplitude modulation (OQAM), SMT allows overlapping the adjacent subcarrier bands, resulting in a flat PSD and hence better utilization of the UWB spectrum.

This paper highlights some of the challenges that are inherent to the UWB communication environment and explores a design based on SMT. Among these challenges, we show that spread spectrum communication is a necessity, due to the low received signal PSD level. Using the anticipated signal to noise ratio (SNR) for a few operating conditions, a few SMT-based transmission schemes are explored. A practical, flexible signaling scheme is selected and a receiver design based on fast convolution filter banks is explored for the chosen scheme. The fast convolution SMT channel equalizer design presented in \cite{6843197} is combined with the multi-channel equalizer design of \cite{730448} to derive a new minimum mean square error (MMSE) channel equalizer for SMT-SS. The equalizer design is further extended to consider narrowband interference (NBI) suppression. Finally, the performance of this design is demonstrated through simulations. It is seen that using the proposed signaling scheme and receiver architecture, FBMC-SS technology provides a promising avenue for addressing the challenges inherent in UWB communications and sensing.

The remainder of this paper is organized as follows. In Section~\ref{sec:uwb_background}, we present a review of the challenges in UWB system design and the previously proposed waveforms. In Section~\ref{sec:fbmcss}, we present the FBMC-SS system proposed in this paper, along with a multi-coding method for maximizing the transmission rate. In Section~\ref{sec:fc}, we present the design of a joint channel equalizer and interference suppression algorithm. In Section~\ref{sec:sims}, we demonstrate the performance of this UWB system design over a simulated UWB multi-path channel with harsh interference. In Section~\ref{sec:conclusion} we conclude.

\section{Challenges in UWB wireless systems} \label{sec:uwb_background}
Early thoughts in the development of UWB systems envisioned data rates as high as $500$~Mbps. Though UWB has potential for these high-rate wireless communications, its adoption by the industry has remained limited. In this section, we discuss several implementation challenges that must be addressed in a successful UWB system design. These challenges include low SNR because of the low transmit power, the presence of high-powered NBI from incumbent radios, and the need for very high-speed analog-to-digital and digital-to-analog convertors (ADC and DAC) because of the very wideband of the UWB signals.

\begin{figure}[tb]
  \centering
  \includegraphics[width=0.95\columnwidth]{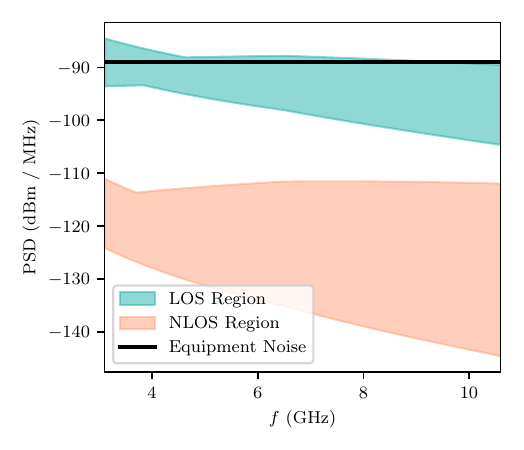}
  \caption{PSD of the equipment noise and the ranges of the received signal PSD as compared the internal noise generated by the radio equipment.}
  \label{fig:SNRs}
\end{figure}

First, following the strategy presented in \cite{UWB2024}, we have implemented a UWB transceiver using Ettus X310 radios. Using the data sheets of this radio equipment and confirming through our own measurements, we assumed this to be a representative PSD a radio equipment. Fig.~\ref{fig:SNRs} presents a diagram showing the PSD of our equipment noise, along with the ranges of the received signal PSD for a set of line-of-sight (LOS) and non-line-of-sight (NLOS) channels based on the path loss equation in the IEEE802.15.4a channel models for an office setting and an industrial site
\cite{channel_model}. The important point to note here is that the received signal is mostly at some negative SNR values. For  NLOS channels, in particular, the SNR range is at $25$~dB or more below the equipment noise. At such a low SNR, even detection of the presence of a data packet may not be a straightforward task.

\begin{figure}[tb]
  \centering
  \includegraphics[width=0.95\columnwidth]{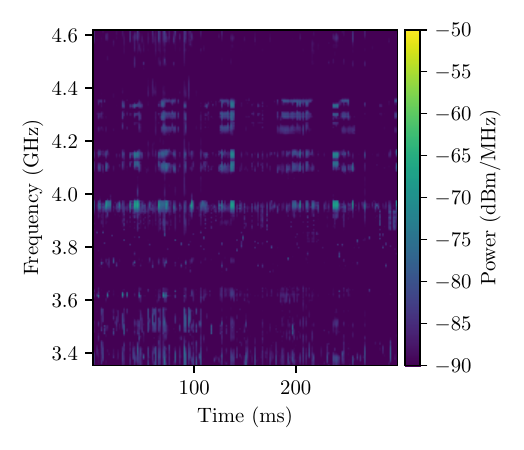}
  \caption{A spectrogram of signal activities near an LTE tower over a bandwidth of $1.28$~GHz.}
  \label{fig:spectroGrm}
\end{figure}

In addition to low SNR, UWB communication systems experience significant NBI caused by the incumbent radios. Significant research into effective NBI suppression algorithms has been done with a good summary provided in \cite{ArslanHüseyin2006NIIi}. A recent study from our group presented in \cite{nelson2024_int_surv} explores the impacts of interference on system outage probability using a vast survey of spectral activity in IEEE802.15.4 operating channels. An example spectrogram capture from this interference study is shown in Fig.~\ref{fig:spectroGrm}. This study found effective suppression of NBI can reduce the probability of system outage  by 1 or 2 orders of magnitude.

Due to the large UWB signal bandwidth, in IR-UWB transmission pulses have a very short duration and, therefore, the receiver can resolve many multi-path components from the operating channel. For IR and DS-UWB, RAKE receivers have been proposed \cite{WinM.Z.1998Otec}, where it is noted that over $50$ RAKE fingers may be required to fully capture  the received signal energy. Alternatively, the MB-OFDM scheme has been suggested as a low complexity method for equalizing the channel, thus, capturing the spread signal energy.

The challenges of interference cancellation and channel equalization are particularly difficult to realize without digital signal processing. Hence, effectively addressing these challenges has proven difficult, partly because of the need for a prohibitively high sampling rate.  In \cite{UWB2024}  an FBMC-SS transceiver architecture that avoids the need for fast ADC and DAC has been proposed. This architecture is further studied in the remaining parts of this paper.

\section{Filter-Bank Multicarrier Spread Spectrum Modulation} \label{sec:fbmcss}
As noted in Section~\ref{sec:intro}, SMT should be used for FBMC modulation in UWB systems. SMT allows for maximization of the transmit power (hence, maximizing the transmission range and/or data rate), while satisfying the spectral mask imposed by the FCC and other regulatory bodies around the world.
SMT makes use of OQAM where the real and imaginary parts of each QAM data symbol are treated as two real-valued symbols and transmitted  at the spacing of $T/2$, where $T$ is the QAM symbol interval. In addition, a $90^\circ$ phase-shift is applied to adjacent (real-valued) data symbols along time and frequency axes. This is to facilitate data recovery at the receiver, making use of what is referred to as real orthogonality in the basis set used in signal synthesis at the transmitter, \cite{OFDMvsFBMC}.

To implement an SMT-based spread spectrum (SMT-SS) each QAM symbol is repeated across a number of $K$ subcarrier bands. This leads to a processing gain $G=10\log_{10}K$~dB. The choice of $K$ depends on the SNR at the receiver input and the desired SNR after despreading. For instance, if there is a LOS channel with an SNR of $-10$~dB at the receiver input and an SNR of $10$~dB is required after despreading, $K$ has to be set equal to $100$. This may be rounded up to $128$ for an effective multi-coding of $7$ bits for each of real and imaginary parts of each QAM symbols, assuming quadrature phase-shift keying (QPSK) signaling. This, for a transmission bandwidth of $1.28$~GHz, leads to a data rate $2\times 7\times \frac{1,280}{128}=140$~Mbps. This data rate will, clearly, vary proportionately with the size of the transmission bandwidth. An example of one such multi-coding scheme in the context of FBMC is discussed in \cite{HaabDavidB2018FBMS}.

As a second example, consider the case of an NLOS channel with an SNR of $-30$~dB at the receiver input, i.e., $20$~dB lower than the case of LOS channel, above. To maintain the same output SNR of $10$~dB, this demands a processing gain which is $100$ times larger, i.e., $10,000$. Rounding this up to the nearest power of two, we get $K=2^{14}$ and a data rate $2\times 14\times \frac{1,280}{2^{14}}\approx 2.2$~Mbps.

Comparing the above two examples, one may notice that while the subcarrier bandwidth in the first example is $10$~MHz, in the second example, it reduces to $10/128=0.078125$~MHz. Accordingly, one may think of the need for a large variation of subcarrier bandwidth, to be able to cater to different needs of the UWB radios. This may be greatly simplified, if one chooses to keep the subcarrier bandwidth fixed to a minimum value and transmit through a set of parallel channels, to adjust the required processing gain and, accordingly, the data rate as explained below.

Consider a case where a total bandwidth of $W$ is divided into $M$ subbands, each with the bandwidth of $B=W/M$. If a given data symbol is spread across all the $M$ subbands, one will end-up with a symbol rate $B$ symbols per second and a processing gain $10\log_{10}M$~dB. Alternatively, if a data symbol sequence is partitioned into $K$ parallel subsequences and each subsequence is spread across $M/K$ subbands, there will be a total symbol rate $KB$ symbols per second and a processing gain $10\log_{10}\frac{M}{K}$~dB. In this way, one can maximize the transmission rate while achieving the required decoding SNR after despreading.

\section{Fast Convolution Filter-Banks:\\ An Effective Tool for Implementation of UWB Transceiver Systems} \label{sec:fc}
Fast convolution (FC) is a classical method for efficient implementation of long finite impulse response (FIR) filters. More recently, the use of FC for implementation of filter-banks has been widely studied and many publications addressing different aspects of such systems have been published, e.g., \cite{RenforsMarkku2014AaDo, Yli-KaakinenJuha2017C8-F, BorgerdingMark2012TOia}. For UWB, fast convolution architectures offer opportunities for low complexity and multi-tap channel equalization \cite{6843197}. In this section, we explore a fast convolution based receiver design and show how a high-performance, low complexity channel equalization scheme can be developed that takes advantage of the diversity introduced by spreading data symbols over a number of subcarriers.

\begin{figure}[tb]
  \centering
  \includegraphics[width=\columnwidth]{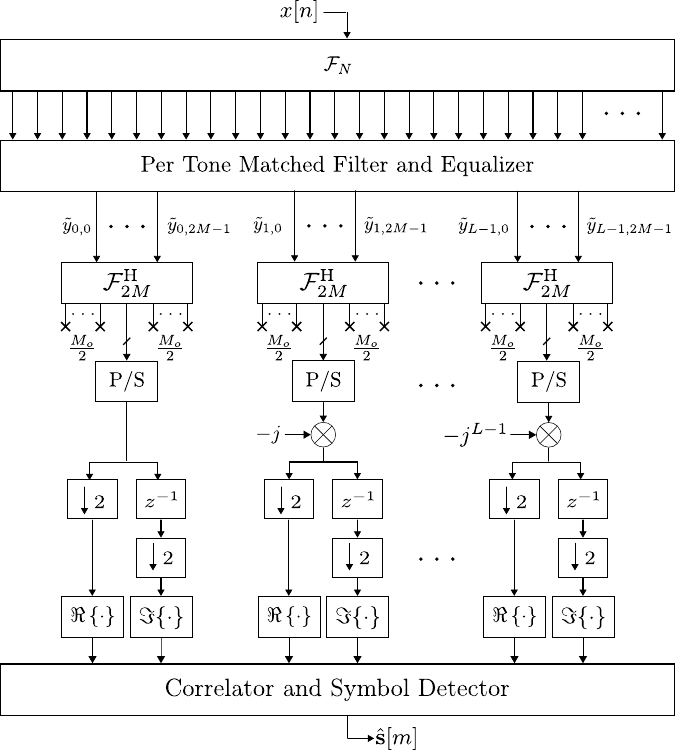}
  \caption{Fast convolution structure for an SMT receiver.}
  \label{fig:FCrx}
\end{figure}

Fig.~\ref{fig:FCrx} shows a diagram of our proposed SMT fast convolution receiver. We design our fast convolution receiver based on the conventional overlap save fast convolution algorithm. First, the input sample stream is partitioned into overlapping blocks of length $N$, where the overlap between adjacent blocks is chosen to be $N_o$. The amount of overlap is chosen so that the portions of the output sample stream impacted by the effects of frequency domain filtering with the prototype filter and the multi-tap equalizer can be removed. The $N$-point DFT is taken of each block, yielding the frequency domain samples $\tilde{x}_k$, for $k=0,\ldots, N-1$. Note that since the adjacent subcarrier bands overlap, the samples at the DFT output share between the adjacent subcarrier bands. For instance, in the left most part of Fig.~\ref{fig:FCrx}, while the first and second blocks of samples are used for data recovery in the first subcarrier band, the second and third blocks of samples are used for data recovery in the second subcarrier band. Moreover, as discussed next, an equalization method that combines signals from different subcarrier bands while taking into account partial band interferers is possible.

Fractionally spaced, multi-tap equalizers for SMT systems have been discussed in many publications, with \cite{BaltarLeonardoGomes2017C1-F} providing a good introduction and summary. These approaches include linear and widely linear MMSE equalization techniques, embedded equalizers in FC-based receivers \cite{6843197}, and multi-stage interference cancellation approaches. In connection with our fast convolution architecture, we choose a frequency domain equalization and interference cancellation scheme that takes advantage of the spread spectrum nature of the proposed SMT-SS design.

Frequency domain equalizers have been discussed in a number of publications, \cite{730448, 6843197, BaltarLeonardoGomes2017C1-F}. They have been shown to offer low complexity per-tone equalization, allowing equalizer weights at different DFT tones to be calculated independently \cite{730448}. Additionally, here, we show taking partial band interference into account in a frequency domain equalizer is also possible and leads to a significant performance improvement.

In designing our channel equalizer and interference suppression scheme, we extend the MMSE designs of \cite{6843197,730448}. Let the $n^{\text{th}}$ frequency domain sample in the $k^{\text{th}}$ subcarrier band corresponding to the $m^{\text{th}}$ data stream be denoted $\tilde{x}^{(m)}_{k,n}$. Also, let $c_{k,n}$ and $h_{k,n}$ denote the corresponding channel and matched filter gains respectively and let $\sigma^2_{k}$ denote the noise plus interference power over the $k^{\text{th}}$ subcarrier band. To arrive at an MMSE equalizer, we first apply a noise whitening step
\begin{equation}\label{eq:normalized tone}
  \tilde{x}^{(m)'}_{k,n} = \frac{\tilde{x}^{(m)}_{k,n}}{\sqrt{\sigma_{k}^2}}.
\end{equation}
Once the noise power has been whitened, the multichannel, frequency domain equalizer of \cite{730448}, combined with the FBMC fast convolution and equalization method of \cite{6843197} leads to
\begin{equation}\label{eq:w'kn}
  \tilde{w}^{(m)'}_{k,n} = \frac{\frac{h_{k,n}c^*_{k,n}}{\sigma_{k}}}{\sum_{l \in \mathcal{S}_m} \left(  \left| \frac{h_{l,n}c_{l,n}}{\sigma_{l}} \right|^2 + \left| \frac{h_{l,n'}c_{l,n'}}{\sigma_{l}} \right|^2 \right) + \frac{1}{\text{SNR}^'_m}},
\end{equation}
where $\mathcal{S}_m$ is the set of all subcarrier indices corresponding to the $m^{\text{th}}$ data stream, $n'$ refers to the tone indices that alias to $n$, and  $\text{SNR}^'_m$ is the averaged SNR across the associated subcarrier bands, given by
\begin{equation}\label{eq:whitened_snr}
  \text{SNR}^'_m = \frac{1}{L}\sum_{l \in \mathcal{S}_m} \text{SNR}_l,
\end{equation}
where $L$ is the number of subcarrier bands covering a data stream.

Making use of \eqref{eq:normalized tone} and \eqref{eq:w'kn}, the equalized signal corresponding to the $m^{\text{th}}$ data stream at tone $n$ is given by
\begin{equation}
  \tilde{y}_{m,n}=\sum_{k\in \mathcal{S}_m }  \tilde{w}^{(m)'}_{k,n}\tilde{x}^{(m)'}_{k,n}.
\end{equation}
Alternatively, this may be rewritten as
\begin{equation}
  \tilde{y}_{m,n}=\sum_{k\in \mathcal{S}_m }  \tilde{w}^{(m)}_{k,n}\tilde{x}^{(m)}_{k,n},
\end{equation}
where $\tilde{x}^{(m)'}_{k,n}$ is replaced by $\tilde{x}^{(m)}_{k,n}$ and, accordingly, we use the equalizer tap weight
\begin{equation}\label{eq:wkn}
  \tilde{w}^{(m)}_{k,n} = \frac{\frac{h_{k,n}c^*_{k,n}}{\sigma_{k}^2}}{\sum_{l \in \mathcal{S}_m} \left(  \left| \frac{h_{l,n}c_{l,n}}{\sigma_{l}} \right|^2 + \left| \frac{h_{l,n'}c_{l,n'}}{\sigma_{l}} \right|^2 \right) + \frac{1}{\text{SNR}^'_m}}.
\end{equation}
Note that there is a factor of $\frac{1}{\sigma_k}$ difference between \eqref{eq:w'kn} and \eqref{eq:wkn}.

The SMT-SS that we have adopted here considers a case where a data stream is transmitted over a number of subcarrier bands. The equalization method presented in \cite{6843197} applies an independent equalizer to each subcarrier band. This reduces \eqref{eq:wkn} to
\begin{equation} \label{eq:w_no_int_suppress}
  \tilde{w}_{k,n} = \frac{\frac{h_{k,n}c^*_{k,n}}{\sigma_{k}^2}}{ \left| \frac{h_{k,n}c_{k,n}}{\sigma_{k}} \right|^2 + \left| \frac{h_{k,n'}c_{k,n'}}{\sigma_{k}} \right|^2  + \frac{1}{\text{SNR}_k}}.
\end{equation}
One may view this as an equalization without interference suppression. This may be understood by recognizing that when the subcarrier bands are equalized independently, each subcarrier band in a data stream is equalized without consideration of the quality of other subcarrier bands that carry the same data. The results from different bands are then simply averaged.   Equation~\eqref{eq:wkn}, on the other hand, may be viewed as a joint equalization and maximum ratio combining across a number of subcarrier bands that are carriers of the same data stream. It gives lower weights to the bands that are highly interfered and/or are affected by a lower channel gain.

In \cite{730448}, the SNR improvement of a multi-channel equalization scheme is compared to the SNR improvement provided by a matched filter. The matched filter provides a performance bound on all equalizers, including maximum likelihood equalizers \cite{730448}. It was seen that as the number of diversity branches grows, the performance of the MMSE equalizer approaches the matched filter bound.

These results hold well for UWB communications because the channel impulse response is characterized by many distinguishable multi-path components. These multi-path components provide many diversity branches, which are then utilized through the tap weight calculations in \eqref{eq:wkn}. Additionally, as seen in \eqref{eq:wkn}, the impacts of NBI are taken into account through the noise-whitening filter. The NBI suppressor scales the channel gain at the impacted frequencies to suppress the interference, effectively equating NBI with a channel fade over the impacted portion of the spectrum. The ``channel fade'' is then addressed through the diversity gain of the remaining subcarrier bands in a data stream.

The SNR improvement provided by this multi-channel equalizer design can also allow for improved sensing and localization performance. In \cite{UWB2024}, it is observed that filter banks provide the same ranging capabilities as IR-UWB and DS-UWB schemes. As discussed above, applying the equalizer and interference suppressor described here allows for an SNR gain that approaches the gain of a matched filter, and hence near optimal SNR, which will improve ranging performance. Development and demonstration of an FB-UWB ranging scheme is beyond the scope of this paper, but will be explored in a future work.

\section{Simulations} \label{sec:sims}
\begin{table}
  \caption{Simulation Configurations}
  \label{tab:sim_cfg}
  \begin{tabular}{|c|c|c|c|c|}
    \hline
    \rowcolor[HTML]{EFEFEF}
    \begin{tabular}[c]{@{}c@{}}Channel\\ Type\end{tabular} & \begin{tabular}[c]{@{}c@{}}Bit Rate \\ (Mbps) \end{tabular} & BW (MHz) & \begin{tabular}[c]{@{}c@{}}Subcarrier\\ BW (KHz)\end{tabular} & \begin{tabular}[c]{@{}c@{}}LDPC Code\\ Rate\end{tabular} \\ \hline
    LOS                                                    & 456.25                                                      & 5120     & 311.9                                                         & 5/7                                                      \\ \hline
    NLOS                                                   & 0.08                                                        & 500      & 243.6                                                         & 1/3                                                      \\ \hline
  \end{tabular}
\end{table}
We demonstrate the performance of the proposed embedded, joint equalizer design through a number of simulations. We compare the performance of the new design with a system that does MMSE equalization independently for each subcarrier band, as proposed in \cite{6843197}.

In our simulations, we consider two parameterizations of the design. One that is tuned to low rate transmissions in NLOS channels and the other that is tuned to high rate transmissions in LOS channels. Key parameters of the two configurations are given in Table~\ref{tab:sim_cfg}. These parameters are chosen based on the anticipated received signal SNR for typical LOS and NLOS scenarios over a range of $3$--$8$ meters. Received signal strength is found using the IEEE802.15.4 channel model and the noise power is chosen based on the noise figure of our development hardware, where it has been found that the SNR in typical LOS scenarios over this range is between $-10$~dB and $+3$~dB. For the case of NLOS channels, the typical received SNR is between $-42$~dB and $-23$~dB.

The simulations use full receiver channel state information (CSI) to compute the equalizer tap weights. Our initial study of channel estimation has revealed that by making use of a long preamble, similar to those recommended in IEEE802.15.4 standard documents, a good estimate of the channel can be obtained, even for NLOS channel where SNR drops to the range of $-40$~dB. These studies will be reported in a future publication once completed.

The simulations use the IEEE802.15.4 industrial LOS and NLOS channel models. The simulations also include harsh primary user interference. The interference environment is characterized based on the results in \cite{nelson2024_int_surv}. It is found that typical interference events range from $5$--$40$~dB above the receiver noise level. For each simulated data block, a number of interfering signals each with a bandwidth of 20 MHz are generated with a random carrier frequency and power spectral density level between $5$ and $40$~dB above the noise power level and added to the received signal. As the system bandwidth grows, the number of potential interfering signals also increases. For this reason, we choose 10 interfering $20$~MHz signals for a high-rate system, and 4 interfering $20$~MHz signals for the low-rate system. The high-rate system has a bandwidth of $5120$~MHz, and the low-rate system bandwidth is equal to $500$~MHz.

\begin{figure}[t]
  \centering
  \includegraphics[width=1\columnwidth]{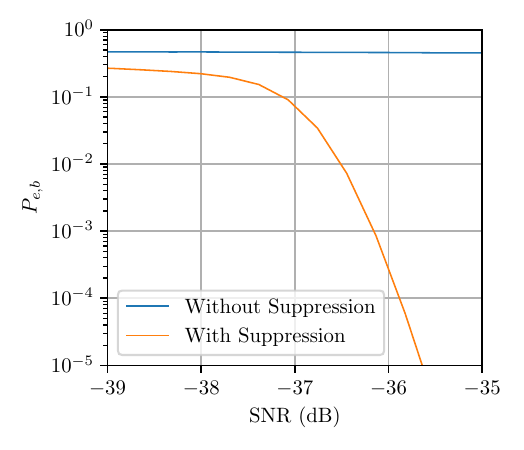}
  \caption{Low-rate BER curves for MMSE equalizers with and without interference suppression.}
  \label{fig:low_rate_ber}
\end{figure}

\begin{figure}[t]
  \centering
  \includegraphics[width=1\columnwidth]{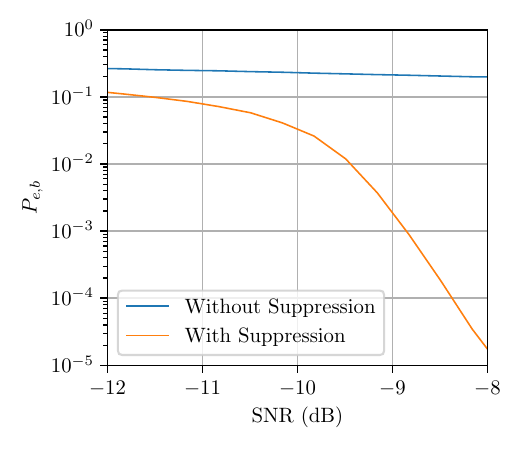}
  \caption{High-rate BER curves for MMSE equalizers with and without interference suppression.}
  \label{fig:high_rate_ber}
\end{figure}

Bit-error-rate (BER) curves for low-rate systems operating over the NLOS office channel are shown in Fig.~\ref{fig:low_rate_ber}. As can be seen, for the proposed joint MMSE equalizer (i.e. equalization and interference suppression), the bit error rate falls off around an SNR of $-36$~dB. Lower SNR performance can be achieved through either a bandwidth expansion or a lower transmission rate. BER curves for the high-rate systems are shown in Fig.~\ref{fig:high_rate_ber}. Here also, one can see the great value of the interference suppression brought in by the proposed multi-channel equalizer \eqref{eq:wkn}.

\section{Experimental Results}
Here, we explore the performance of the proposed joint equalization and interference suppression by applying interference plus noise captured by a UWB hardware to a simulated UWB signal and looking at the BER results. The details of UWB hardware along with the details of some of our captured data can be found in \cite{nelson2024_int_surv}. We use the IEEE802.15.4 channel model to synthesize a UWB signal and adjust the received power-level against our equipment noise to arrive at different choices of SNR. That is, the added interference from the environment is not accounted for in the presented SNR values.  The joint equalization and interference suppression takes the necessary steps for effective extraction of the transmitted bits.

\begin{figure}[t]
  \centering
  \includegraphics[width=1\columnwidth]{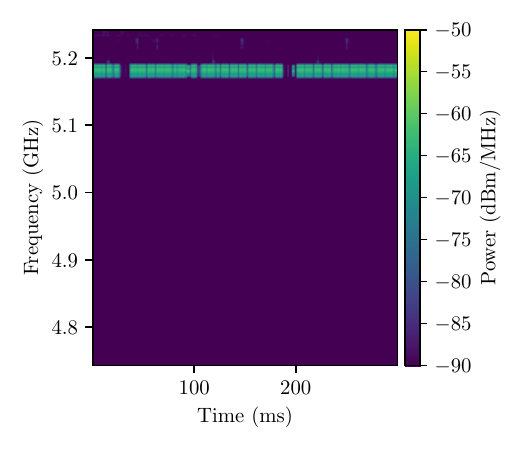}
  \caption{Sample spectrogram of the interference environment used in the BER curve of Fig.~\ref{fig:exp_chan3}.}
  \label{fig:spec_icc_chan3}
\end{figure}

\begin{figure}[t]
  \centering
  \includegraphics[width=1\columnwidth]{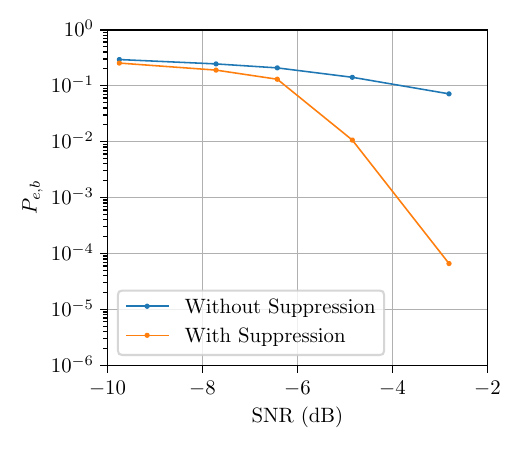}
  \caption{BER curves based on experimental data captures for MMSE equalizers with and without interference suppression.}
  \label{fig:exp_chan3}
\end{figure}

Fig.~\ref{fig:exp_chan3} presents our results for the case of a LOS office channel model.  The system is configured to have a bit rate of $124.75$~Mbps, a bandwidth of $500$~MHz, a subcarrier spacing of $243.6$~kHz, and an LDPC code rate of $1/2$. The transmission is over Channel $3$ of IEEE802.15.4 standard, where the center frequency is $4992.8$~MHz. As seen, the proposed joint equalization and interference suppression results in a significant improvement  when compared to a receiver that is not equipped with interference suppression. Looking more into our interference environment, we found that in most cases there exist one or more high-power NBIs in the band of transmission. Such interferers drop the signal to interference plus noise ratio (SINR) to a level that cannot support the simulated data rate.  A sample spectrogram of the  interference that we have used in generating Fig.~\ref{fig:exp_chan3} is shown in Fig.~\ref{fig:spec_icc_chan3}.

\section{Conclusion} \label{sec:conclusion}
It was noted that  ultra-wideband (UWB) systems, used for  communications, sensing, and localization, face a number of challenges whose effective solutions are only possible through use of advanced signal processing methods. Following a recent proposal \cite{UWB2024}, this paper presented a detailed study of filter-bank multicarrier spread spectrum (FBMC-SS) as an effective solution for implementing UWB systems. It was noted that FBMC-SS facilitates (i) straightforward adoption of the spectral masks mandated by different standard bodies around the world; (ii) partitioning the UWB signal in to a number of medium size bands that may be processed in parallel at both the transmitter receive sides, hence, avoiding the need for very high-speed DAC and ADC components; and (iii)  blindly removing the narrow-band interferers coming from the incumbent radios.

To demonstrate the effectiveness of these solutions, we implemented a UWB system using of a set of four Ettus X310 radios. This led to a  total bandwidth of $1.28$~GHz, consisting of eight $160$~MHz subband sections. We also presented a transceiver FBMC-SS design based on a fast convolution filter-bank structure. We developed a joint channel equalization and blind interference suppression method. Finally, through simulations, the performance of the proposed structure was verified and compared with prior approaches. It was confirmed that, to address the unique challenges of UWB in harsh environments, a joint equalization and interference cancellation algorithm is necessary.

\section*{Acknowledgements}
This research made use of Idaho National Laboratory's High Performance Computing systems located at the Collaborative Computing Center and supported by the Office of Nuclear Energy of the U.S. Department of Energy and the Nuclear Science User Facilities under Contract No. DE-AC07-05ID14517.

\thanks{This manuscript has in part been authored by Battelle Energy Alliance, LLC under Contract No. DE-AC07-05ID14517 with the U.S. Department of Energy. The United States Government retains and the publisher, by accepting the paper for publication, acknowledges that the United States Government retains a nonexclusive, paid-up, irrevocable, world-wide license to publish or reproduce the published form of this manuscript, or allow others to do so, for United States Government purposes. STI Number: INL/CON-24-81456.}

\bibliographystyle{IEEEtran}

\bibliography{IEEEabrv,ref, uwb_link_budget}

\end{document}